# Dynamics of Dissociative Electron Attachment to Aliphatic Thiols

Sukanta Das, and Vaibhav S. Prabhudesai

*Tata Institute of Fundamental Research, Colaba, Mumbai 400005 India*

**Abstract:**

Dissociative electron attachment (DEA) shows functional group-dependent site selectivity in the $H^-$ ion channel. In this context, the thiol functional group has yet to be studied in great detail, although this functional group carries importance in radiation damage studies where the low-energy secondary electrons are known to induce damage through the DEA process. We report detailed measurements of absolute cross-sections and momentum images of various anion fragments formed in the DEA process in simple aliphatic thiols. We also compare the observed dynamics with that reported earlier in hydrogen sulphide, the precursor molecule for this functional group and also with aliphatic alcohols. Our findings show substantial resemblance in the underlying dynamics in these compounds and point to a possible generalisation of these features in the DEA to thiols. We also identify various pathways that contribute to the $S^-$ and $SH^-$ channels.

# 1. Introduction:

Low-energy electron collisions with organic molecules have attracted much attention lately after the discovery of their direct[1] and indirect[2] role in the single and double-strand breaks in DNA that may lead to radiation damage to living cells. Apart from the low-energy electrons formed in the energy range of $0 - 20$ eV due to the high-energy radiation interaction with the organic matter, the radicals formed during these interactions also cause considerable damage to the living cells[3]. In this context, many efforts have been made to understand the interaction of low-energy electrons with organic molecules. The functional group-dependent site-selectivity observed in the dissociative electron attachment (DEA) to organic molecules has unravelled new facets of the low-energy electron molecule interaction[4]. The molecules containing a hydroxyl group are particularly interesting and have been extensively studied in the past[5–8]. Thiol is one of the important molecular functional groups responsible for many processes inside the biological system. The molecules that have a thiol group primarily reside in the cell and play a vital role in different cellular functions like glycolysis, mitochondrial energy production, and stimulation of glucose metabolism[9]. Thiols are one of the main protective mechanisms against oxidative stress[10]. It also plays a role in enzymatic reactions, detoxification, and antioxidant protection in the body[11]. However, DEA to thiol compounds has not been explored extensively, apart from one report on ethanethiol[12].

On the other hand, a comparison of DEA to water and hydrogen sulphide provides an interesting case to understand the intricate molecular dynamics followed by the negative ion resonances containing the OH and SH groups. For example, the two molecules show distinct resonances dissociating into the $H^-$ channel in DEA. These resonances are at 6.5, 8.5 and 12 eV in water[13] and around 5.2, 7.5 and 9.6 eV in hydrogen sulphide[14]. Both these molecules belong to the $C_{2v}$ symmetry group. As O and S atoms belong to the same group in the periodic table, these molecules are expected to show similar structures and dynamics. On the other hand, the O atom is more electronegative than the S atom, making it a stronger hydrogen bond participant. The negative ion resonances corresponding to the three peaks observed in the $H^-$ channel in DEA are $^2B_1$, $^2A_1$ and $^2B_2$ states which are core excited resonances. However, there are distinct differences in their dynamics as observed in the momentum imaging[14,15]. The $H^-$ channel at the first resonance in water shows substantial internal excitation of the OH fragment, whereas in hydrogen sulphide, the SH fragment shows very little internal excitation. The angular distribution of the $H^-$ channel from the second resonance in water shows significant internal motion before dissociation, making it deviate from the expected angular distribution from the axial recoil motion[16,17]. In contrast, in hydrogen sulphide, the angular distribution at the second resonance is in accordance with the axial recoil motion[14]. More interestingly, the $OH^-$ ions are reported to be observed experimentally by Fedor *et al.*[18] as the direct dissociation product from all three resonances, while theoretical calculations do not find this channel as a direct product in these resonances and attribute it to nonadiabatic effects[19]. On the other hand, in DEA to hydrogen sulphide, $SH^-$ ions are reported at 2.4 eV[14,20]. Interestingly, both simple alcohols[21] and thiols[12] have shown the production of $OH^-$ and $SH^-$ ions, respectively. Moreover, the systematic investigations in simple organic alcohols have shown the presence of an $H^-$ channel at around 6.5 eV that arises from the hydroxyl group and shows strong functional group dependence[4]. However, at 6.5 eV resonance, using the momentum imaging technique, it was found that the corresponding angular distribution of the $H^-$ channel is substantially different from that reported for $H^-$ from water at 6.5 eV despite a similar resonance state playing a role in the reaction. The difference in the angular distribution has been explained as due to the torsional modes of vibrations active at room temperature[6,8]. These vibrational excitations have a substantial influence on the angular distribution. In light of these observations, it would be interesting to explore the similar functional group dependence of the DEA process in thiol groups.

Ibănescu and Allan have reported $C_2H_5S^-$, $C_2H_4S^-$, $SH^-$, and $S^-$ fragments from DEA to ethanethiol[12]. $C_2H_4S^-$ shows two peaks at 0.61eV and 1.66eV, $CH_3CH_2S^-$ shows a peak at 1.83eV, and all these peaks are assigned as shape resonances. While $SH^-$ peaks at 8.7eV and $S^-$ at 8.1eV, they are inferred to arise from core excited Feshbach resonances.

Here, we report the detailed investigation of the DEA process in simple thiols, namely ethanethiol and 1-butanethiol. Taking the cue from the hydrogen sulphide, we expect the thiol group-containing compounds to follow the axial recoil motion, throwing more light on the underlying DEA dynamics. In this work, we have obtained the resonance position of different fragments from ethanethiol and 1-butanethiol and, estimated their kinetic energy (KE) and angular distribution from the momentum images taken using the velocity slice imaging (VSI) technique and compared them with the $H_2S$ and ethanol, especially at 6.5 eV channel of ethanol. We have also found an energetic $S^-/SH^-$ channel. This channel is important in atmospheric chemistry, particularly in its sulphur budget from organic mercaptans[22].

## 2. Experimental Setup:

The experimental setup is described in detail elsewhere[23], and here we provide its brief description. Magnetically collimated pulsed electron beam produced from a home-built thermionic emission-based electron gun is made to cross the molecular beam at a right angle. The gaseous molecular beam is produced using a capillary array connected to the glass bulb that contains the pure sample of the target molecule in liquid form. The bulb is evacuated before introducing the liquid and pumped using a dry pump until the sample volume is reduced to 1/3rd. The pulsed electron beam (width 100 ns and repetition rate of 10 kHz) is collected by a Faraday cup mounted coaxial to the electron gun but on the opposite side of the interaction volume. The electron current is measured using an electrometer (Kethley 6512). The molecular beam is introduced in the interaction region along the axis of the VSI spectrometer. The VSI spectrometer comprises an interaction region flanked by two electrodes, namely pusher and puller, with the latter having a molybdenum wire mesh of 64% transmission spot-welded across its central aperture of diameter 50 mm. The pusher-puller distance is kept at 20 mm. The puller electrode is followed by a four-element electrostatic lens assembly and a short flight tube of 25 mm in length. The two-dimensional position-sensitive detector is comprised of a pair of micro-channel plates (MCP) in the chevron configuration and a phosphor screen with a P47 scintillator and is mounted co-axially with the flight tube on its other end. The ions generated in the interaction region are pushed into the direction of the detector using a square pulsed voltage (-100V amplitude and 1-microsecond duration) applied to the pusher electrode with the puller electrode kept at zero potential. The extraction pulse is synchronised with the electron pulse using a delay generator, maintaining a delay of 50 ns. The signal due to ions is obtained for the time-of-flight measurement using a coupling circuit connected to the MCP back. The ion-yield curves so obtained are put on the absolute scale using the relative flow technique[24], where the DEA to oxygen at 6.5 eV is used as a standard gas to obtain the absolute cross-section of the process[25]. After identifying the ion, the position information of a particular ion hit is obtained by recording the light signal from the phosphor screen using a CCD camera. To get a time slice of the Newton sphere of a particular ion species, the MCP detector is synchronously pulsed (pulse width of 80 ns) using a high-voltage switch. This pulsing is synchronised with the electron pulse with an appropriate delay to coincide with the centre of the time-of-light peak of that particular ion species. The ion hit distribution obtained using a CCD camera is added frame-by-frame to obtain their final distribution. As for the effusive molecular target, the background gas contributes substantially to the detected signal, especially when the projectile electron beam is not focused in the interaction region, and this background signal produces substantial distortion in the image. The imaging condition is optimised to minimise this distortion, and we have subtracted the background signal from the crossed-beam signal with appropriate normalisation with the electron current to avoid any additional artefacts[26].

## 3.    Result and Discussion:

The DEA measurements on ethanethiol and 1-butanethiol show $H^-$ as the most dominant channel apart from the $S^-/SH^-$ and $(M-1)^-$ ions. The mass spectrometer could not resolve the $S^-$ and $SH^-$ ions due to poor resolution. Similarly, the $(M-1)^-$ ions may also have a contribution from $(M-2)^-$. However, the mass spectrometer used in these measurements could not distinguish between the two. The absolute cross-section curves for various ions as a function of electron energy starting from 1 eV are shown in Figure 1. The $H^-$ channel from both molecules shows three peaks around 5eV, 7eV, and 8eV. The first two peak positions are close to those obtained from the hydrogen sulphide, the precursor molecule of the SH functional group. As can be seen from Figure 1, $H^-$ is the most dominant channel with a maximum cross-section of $4 \times 10^{-19}$ cm$^2$ at 5 eV for ethanethiol and $8.7 \times 10^{-20}$ cm$^2$ at 7.8 eV for 1-butanethiol. All the cross-section values have an uncertainty of about 15%, with the maximum contribution from the reference gas cross-section value.

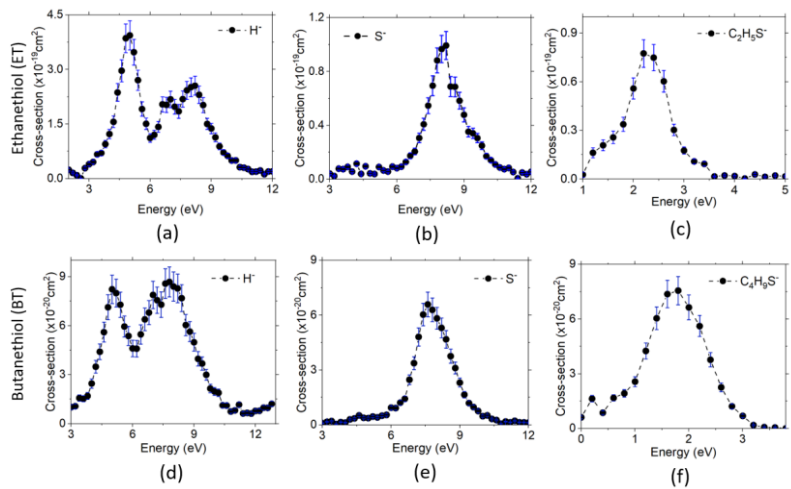

Figure 1: The ion-yield curve obtained for the DEA process and put on the absolute cross-sections scale for the (a) $H^-$, (b) $S^-/SH^-$, and (c) ($C_2H_5S^- + C_2H_4S^-$) channels from ethanethiol and (d) $H^-$, (e) $S^-/SH^-$, and (f) ($C_4H_9S^- + C_4H_8S^-$) channels from 1-butanethiol as a function electron energies.

Unlike hydrogen sulphide, both the thiol molecules show no resonance in the $S^-/SH^-$ channels at low energy. These channels peak around 8.2 eV for ethanethiol and 7.6 eV for 1-butanethiol, with the absolute cross-section comparable with the $H^-$ channel from the respective molecules. The $(M-1)^-$ ion signal peaks around 2.2 eV and 1.8 eV for ethanethiol and butanethiol, respectively. The absolute cross-section values estimated for all these channels at various resonances are given in Table 1.

Table 1: Absolute cross-section of various DEA channels from ethanethiol and butanethiol.

| Ion species | Ethanethiol | | Butanethiol | |
|---|---|---|---|---|
| | Electron energy (eV) | Absolute Cross-section ($\times 10^{-19}$ cm$^2$) | Electron energy (eV) | Absolute Cross-section ($\times 10^{-20}$ cm$^2$) |
| $H^-$ | 5 | $3.93 \pm 0.40$ | 5 | $8.23 \pm 0.87$ |
| | 7 | $2.17 \pm 0.22$ | 7 | $7.88 \pm 0.83$ |
| | 8.2 | $2.55 \pm 0.26$ | 7.8 | $8.67 \pm 0.90$ |
| $S^- + SH^-$ | 8.2 | $0.96 \pm 0.10$ | 7.6 | $6.59 \pm 0.67$ |
| $(M-1)^-$ | 2.2 | $0.77 \pm 0.08$ | 1.8 | $7.56 \pm 0.77$ |

We have obtained the momentum images of all these channels at various electron energies using the VSI technique. We deduced the kinetic energy and angular distributions for these channels from the offline data analysis of these momentum images. Below, we provide details of these distributions for each channel and describe the inferred related molecular dynamics.

### 3.1 $H^-$ ions

This channel shows three peaks at 5, 7 and 8 eV. The momentum images obtained for this channel at various electron energies for both molecules are shown in Figure 2.

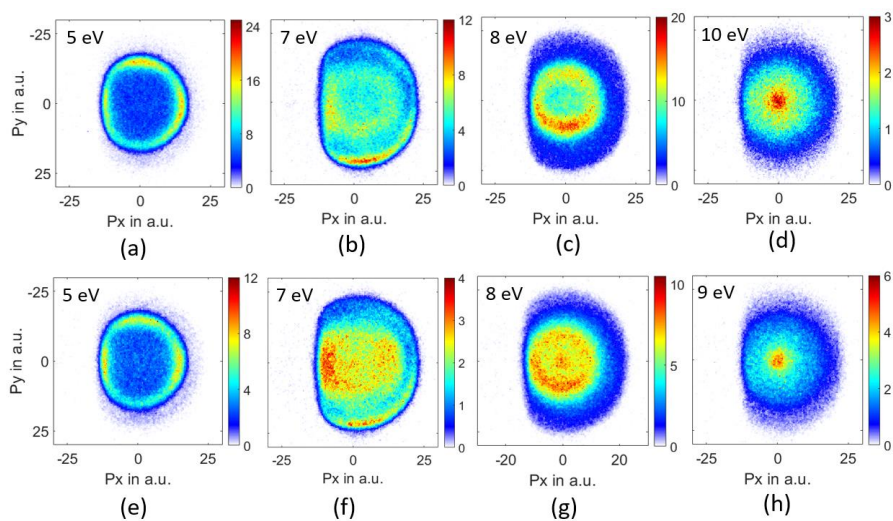

Figure 2. The momentum images for the H$^-$ ions at (a) 5 eV, (b) 7 eV, (c) 8 eV and (d) 10 eV from ethanethiol and (e) 5 eV, (f) 7 eV, (g) 8 eV and (h) 9 eV electron energy from 1-butanethiol. The direction of the electron beam is from top to bottom.

Due to the presence of an electron-beam-collimating transverse magnetic field, the ion images are deflected away from the axis of the spectrometer. This deviation is maximum for the lightest H$^-$ ions. Hence, the momentum image of these ions shows maximum distortion on one side[27]. As the DEA process has azimuthal symmetry about the electron beam, the momentum image is expected to show the left-right symmetry. For the kinetic energy and angular distribution estimation, we have used only the right half of the image, measured close to the detector axis and is distortion-free. The kinetic energy distributions obtained for this channel from both the molecules at various electron energies are shown in Figure 3.

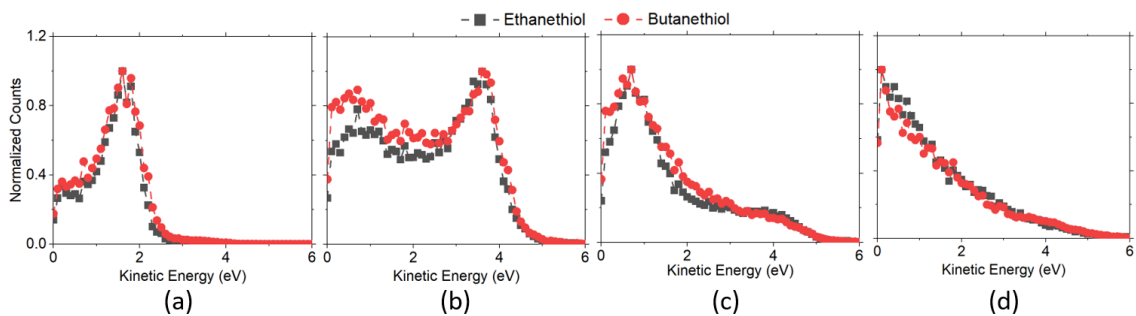

Figure 3: Kinetic energy distribution for the H$^-$ ions from ethanethiol (-■-) and 1-butanethiol (-●-) at (a) 5 eV, (b) 7 eV, and (c) 8 eV electron energy. (d) The kinetic energy distribution obtained for ethanethiol (-■-) at 10 eV and 1-butanethiol (-●-) at 9 eV electron energy.

As mentioned earlier, the first two peaks in the ion-yield curve, at 5 eV and 7 eV, are consistent with those observed in the hydrogen sulphide, the precursor molecule of the thiol (-SH) functional group. Based on the earlier work on the functional group-dependent site-selective fragmentation, we conclude that these two peaks arise exclusively from the S-H site of the molecules.

The threshold for obtaining this ion in a two-body break up from the parent molecule can be determined using the heat of formation for various components. Taking the heat of formation for $C_2H_5SH$ as -46 kJ/mol[28], $C_2H_5S$ as 104 kJ/mol[29] and H as 218 kJ/mol and taking the electron affinity of H as 73 kJ/mol, we get 295 kJ/mol or 2.97eV as the minimum energy required to obtain this ion in DEA to ethanethiol. For 1-butanethiol with the heat of formation of 1-$C_4H_9SH$ as -87.9 kJ/mol and that for 1-$C_4H_9S$ as 54.4 kJ/mol[30], we get the threshold energy for the $H^-$ channel as 2.98 eV. Hence, the excess energy in the system would be 2.03 eV (2.02 eV) for ethanethiol (1-butanethiol). As $H^-$ is the lightest fragment, in the axial recoil motion, most of the excess energy will show up as its kinetic energy. The kinetic energy distribution for this channel at 5 eV peaks around 1.75 eV for both molecules, indicating fast dissociation. Although the maximum expected kinetic energy for 5 eV electron energy is 2 eV, the kinetic energy distribution shows a spread up to 2.5 eV. We attribute it to the electron beam's poor energy resolution (about 0.8 eV). For the 7 eV resonance as well, the kinetic energy distribution has a peak between 3.5 and 3.75 eV, which is consistent with the fast two-body breakup scenario. The angular distribution obtained around this peak in the kinetic energy distribution is shown in Figure 4 for both the molecules at 5 eV and 7 eV electron energies.

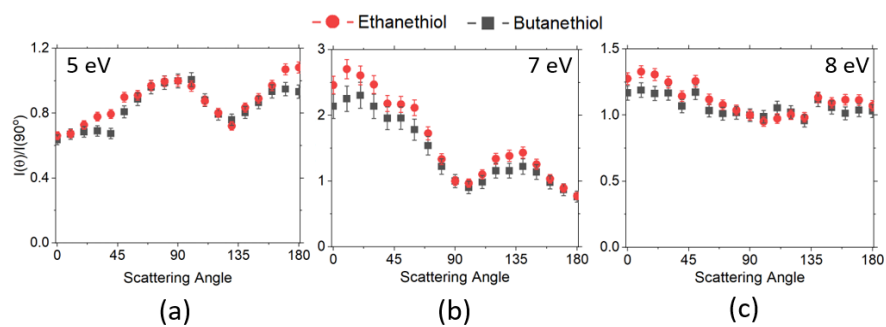

Figure 4: The angular distribution of $H^-$ ions from ethanethiol (-●-) and 1-butanethiol (-■-) at (a) 5 eV (kinetic energy range ), (b) 7 eV (kinetic energy range 3 to 4.2 eV )and (c) 8 eV (kinetic energy range ) peaks in the ion-yield curve.

The angular distribution at 5 eV has two distinct features (Figure 4(a)). The angular distribution peaks around $90^0$-$100^0$, dips close to $130^0$ and then again rises close to $180^0$. The $100^0$ peak matches fairly well with the angular distribution obtained for $H_2S$ at 5 eV[14] with one significant difference. The angular distribution for $H_2S$ becomes extremely small in the forward and backward direction, which is not the case with either of the thiol molecules. However, in this context, it resembles the angular distribution obtained for the first resonance in this channel in ethanol[8]. In hydrogen sulphide, similar to water, the 5.5 eV resonance is understood to be the core excited resonance, with a significant contribution coming from the excitation of the lone pair of electrons from the sulphur atom to the $6a_1$ orbital with the incoming electron getting captured in it. This makes the resonant state a $^2B_1$ state, and the angular distribution resembles that of water with zero intensity in the forward and backward directions[31]. However, the presence of the DEA signal in the forward and backward direction in alcohols has been identified as the contribution of the torsionally excited molecules. We also attribute the signal in the forward and backward direction to the torsionally excited thiol molecules. The major difference between the angular distribution from ethanol and ethanethiol is that the contribution in the backward direction in the latter is far lower than that in the former[8]. Moreover, the signal in the forward direction shows the reverse trend. This points to the possibility of differences in the dynamics of the torsionally excited molecules in both species on electron attachment. Interestingly, the spread in the kinetic energy distribution in both ethanol and ethanethiol is comparable, unlike in water and hydrogen sulphide, showing similarity in the overall dynamics followed by the anions formed by electron attachment to the ground state molecules. Another point to note is that the angular distribution observed in 1-butanethiol is similar to that observed for ethanethiol. The increase observed in the signal in the backward direction from methanol[6] to ethanol[8] is not observed in the ethanethiol and butanethiol, although for these molecules, the number of torsional degrees of freedom increases, and their frequency decreases.

Compared with the first resonance, the second resonance in ethanethiol shows a substantially different angular distribution compared to ethanol. The angular distribution obtained at 7 eV for the $H^-$ channel with kinetic energy in the range of 3 to 4.2 eV (Figure 4(b)) resembles that obtained for the second resonance in this channel in hydrogen sulphide. In water, this resonance shows substantial internal dynamics[16,17]. Similar internal dynamics also result in the second resonance in ethanol, which is reflected in a more smeared angular distribution with reduced anisotropy[8]. However, hydrogen sulphide shows very little internal dynamics, resulting in the angular distribution following the expected pattern from the axial recoil approximation[14]. Comparing the angular distribution for the second resonance in ethanethiol with ethanol and hydrogen sulphide, we conclude that this resonance also follows the axil recoil motion in the dissociation process. For 1-butanethiol, this resonance shows similar dynamics to that of ethanethiol.

The momentum image at the third peak at around 8 eV shows a ring along with a considerable signal at lower energies (Figure 2 (c)) for ethanthiol. This low-energy feature clearly appears as a distinct blob in the case of 1-butanethiol (Figure 2 (g)). The 8 eV peak is broad in the ion-yield curve, and its contribution at lower energies can be seen in the momentum image at 7 eV as the inner ring. The images taken at higher energies (10 eV for ethanethiol and 9 eV for 1-butanethiol) show a clear, low-energy feature. This indicates the presence of another resonance in this energy range. On comparing with the momentum images obtained for methanol[32] and ethanol[8] around these energies, we identify this feature arising from the C-H site in the molecule. This site breakage results from a many-body break-up mechanism, leaving very little energy as the kinetic energy of the fragment. It is also known that the resonance peak in the $H^-$ channel from alkanes shifts to the lower electron energies with the increasing number of carbon atoms in the moelcule[33]. This is consistent with the observation that this low-energy feature appears prominently in 1-butanethiol compared to ethanethiol in the momentum image at 8 eV.

The higher kinetic energy feature (peaking between 0.5 and 0.75 eV) shows an almost isotropic distribution from 1-butanethiol, whereas ethanethiol shows slightly higher intensity in the forward direction compared to the backward direction. This feature starts appearing in the momentum image taken at 7 eV (Figure 2 (b)). The angular distribution obtained for this ring in the kinetic energy range 0.45 – 0.9 eV (0.35 – 0.8 eV) for ethanethiol (1-butanethiol) is shown in Figure 4 (c). The kinetic energy spread and the angular distribution suggest that the source of this channel is a two-body dissociation. However, for 8 eV electron energy, the excess energy for the two-body breakage path with both the fragments in the ground state is almost 5 eV. The observed kinetic energy range from 0 to 2eV implies that the excess energy with the molecular fragment ($C_2H_5S$) is in the range of 3 to 5 eV. However, this fragment has an absorption band with the band origin at around 2.9 eV. The corresponding excited state is identified as $B\,^2A''$ state[34]. This implies that the observed $H^-$ channel arises from the dissociation path

$$e^- + C_2H_5SH \rightarrow (C_2H_5SH^-)^* \rightarrow C_2H_5S\,(B\,^2A'') + H^-$$

(1)

where the molecular fragment $C_2H_5S$ is electronically excited. Although no information is available in the literature about the excited states of 1-$C_4H_9S$ species, a similar momentum image obtained for 1-butanethiol indicates the existence of such an excited state similar to that of ethanetio radical. A more isotropic angular distribution of this channel for 1-butanethiol (Figure 4 (c)) compared to ethanethiol indicates a comparatively slower initial motion of the dissociating species.

## 3.2  $S^-$ and $SH^-$ ions:

The experimental setup used in these measurements cannot separate the ions from the $S^-$ and $SH-$ channels. The combined ion yield curve for these channels shows a single peak at 8.2 eV (7.6 eV) for ethanethiol (1-butanethiol) (Figure 1 (b) and (e)). For the VSI condition used for the measurements, the time-of-flight signal for the two ions was expected to appear within the width of the time slice (80 ns). Hence, in the momentum images, we cannot distinguish between the two ions. According to Ibănescu and Allan, both ions have equal intensity and peak at 8.1 eV with an additional peak at 8.7 eV in the $SH^-$ channel[12]. The mass spectrometric measurements in the modified spectrometer with a relatively better mass resolution clearly show both the ions. Their relative strengths as a function of electron energy show that the 8.2 eV peak observed in the combined ion yield curve is dominated by the $S^-$ ions along the rising edge of the peak. At the peak the two ions have almost equal contributions and on the trailing edge the signal is dominated by the $SH^-$ ions. We have obtained the momentum images of these ions at various electron energies across the 8.2 eV peak for ethanethiol. These momentum images are shown in Figure 5 along with the observed mass spectrometric identification of the ion species.

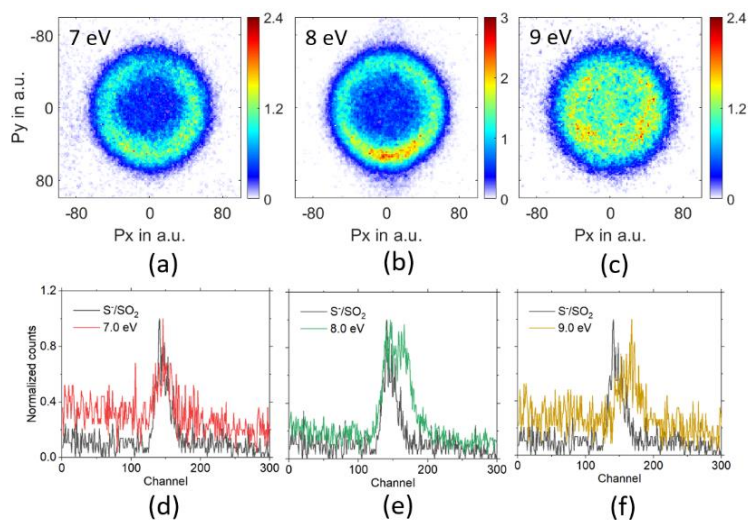

Figure 5: Momentum image of $S^-/SH^-$ at (a) 7 eV, (b) 8 eV and (c) 9 eV electron energy. The direction of the electron beam is from top to bottom. (d) to (f) are the mass spectra obtained for the corresponding electron energies in this mass range using the mass spectrometer with relatively better mass resolution.

For 7 eV electron energy, the momentum image shows a ring with an anisotropy, which appears to be on top of an isotropic distribution. The anisotropy increases at 8 eV. However, the ring becomes slightly narrower and intense. At 9 eV, the inner part of the image gets filled showing the production of low energy ions. We obtained the kinetic energy distribution from these images, and the results are shown in Figure 6 (a).

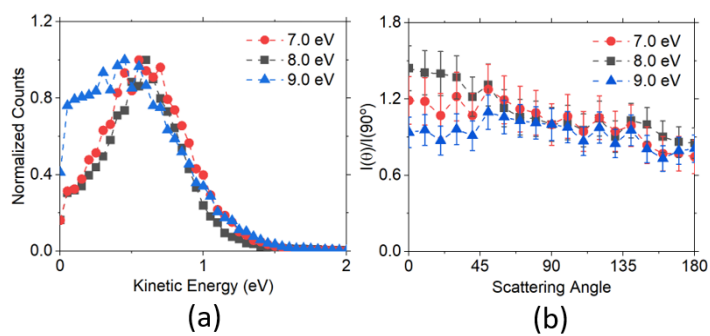

Figure 6: (a) Kinetic energy distribution of $S^-/SH^-$ ions obtained from the momentum images taken at 7 eV, 8 eV and 9 eV for ethanethiol. (b) Corresponding angular distribution of these ions taken in the kinetic energy range 0.4 eV to 0.85 eV.

The kinetic energy distribution peaks between 0.5 and 0.75 eV for both 7 eV and 8 eV electron energies. The ion signal at 7 eV is dominated by the $S^-$ ions. However, this ion is observed with kinetic energy as well as anisotropy. Typically, there are three possibilities of getting $S^-$ ion in DEA o ethanethiol. The first one being from the many body dissociation, where multiple bonds are simultaneously broken.

$$e^- + C_2H_5SH \rightarrow (C_2H_5SH^-)^* \rightarrow C_2H_5 + S^- + H \qquad (2)$$

The thermodynamic threshold for this channel would be 4.77 eV ($\Delta H_f(C_2H_5) = 119$ kJ/mol, $\Delta H_f(S) = 277$ kJ/mol, E.A.(S) = 2.07 eV)[28]. In this case, one of the fragments, H atom, is likely to take away most of the excess energy in the system as its kinetic energy. Interestingly, similar channel is observed in hydrogen sulphide at this energy[13]. Understandably, the $S^-$ channel there shows very low kinetic energy release as the other fragments are H atoms. Hence, this channel may contribute to the low energy part of the image.

The other mode would be a two body break up after the rearrangement of the H atom from the SH site. This channel would be slow due to the H migration in the molecule.

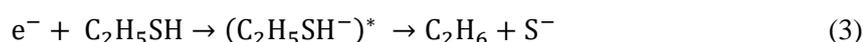
$$e^- + C_2H_5SH \rightarrow (C_2H_5SH^-)^* \rightarrow C_2H_6 + S^- \qquad (3)$$

The thermodynamic threshold for this channel is 0.47 eV ($\Delta H_f(C_2H_6) = -84$ kJ/mol, $\Delta H_f(S) = 277$ kJ/mol, E.A.(S) = 2.07 eV)[28]. Based on the amount of excess energy available in the system, this channel may contribute to the higher energy ring. However, it requires substantial internal rearrangement and it is unlikely to show the anisotropy as seen in the momentum image.

The third channel of the formation of $S^-$ ions is via a sequentially dissociation, i.e. the formation of $SH^-$ ions, followed by its further dissociation into $S^-$ ions.

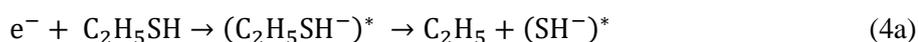
$$e^- + C_2H_5SH \rightarrow (C_2H_5SH^-)^* \rightarrow C_2H_5 + (SH^-)^* \qquad (4a)$$

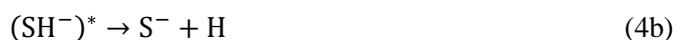
$$(SH^-)^* \rightarrow S^- + H \qquad (4b)$$

This is possible if the $C_2H_5$ radical is formed in the electronic ground state and $SH^-$ ions are formed in their dissociating part of an excited state. Based on the heat of formation of fragments ($\Delta H_f(SH) = 139$ kJ/mol) and electron affinity of SH (EA(SH) = 2.3 eV)[28], we estimate the dissociation energy of the $SH^-$ ion as 3.92 eV. This implies the minimum energy required to obtain the $S^-$ ions by this channel is 4.77 eV. However, the first step (4a) would have at most 2.23 eV as the kinetic energy for 7 eV electron energy. This implies that the maximum kinetic energy of the $(SH^-)^*$ would be 1.04 eV. In this case, the $S^-$ ions would have very little kinetic energy in the centre of mass frame of the $SH^-$ ions. Corresponding linear momentum of the $SH^-$ ion would be passed on to the $S^-$ ion formed in the centre of mass frame of the $SH^-$ ion. This makes the maximum kinetic energy of the $S^-$ ion in this channel to be 1.07 eV. This is consistent with the observed kinetic energy distribution of the ring. Moreover, the angular distribution of the $S^-$ ions would be that of the $SH^-$ ions, explaining the observed anisotropy. In the absence of the $SH^-$ signal at 7 eV, we attribute the observed $S^-$ signal to this channel.

At 8 eV, we also see the $SH^-$ ions with almost same intensity as that of $S^-$ ions. However, we do not see substantial change in the kinetic energy distribution, and the anisotropy increases compared to 7 eV image.

The SH⁻ channel in ethanethiol can result from either a two-body or a three-body break-up. The possible reaction paths are

$$e^- + C_2H_5SH \rightarrow (C_2H_5SH^-)^* \rightarrow C_2H_5 + SH^- \quad (5)$$

$$e^- + C_2H_5SH \rightarrow (C_2H_5SH^-)^* \rightarrow CH_3 + CH_2 + SH^- \quad (6)$$

The threshold for the path (5) is 0.85 eV. Similarly, for path (6) ($\Delta H_f(CH_3) = 145$ kJ/mol, $\Delta H_f(CH_2) = 386$ kJ/mol)[28], we have obtained the threshold as 5.13 eV. There can be other three-body break-up paths where one of the fragments in the concerted break-up can be an H atom. However, in such cases, most of the excess energy will be carried by the H atom, the lightest fragment. As the kinetic energy distribution peaks at about 0.75 eV with a well-defined relatively narrow spread, we rule out any such concerted break-up mechanism resulting in this fragment at this energy. Moreover, from path (6), the kinetic energy distribution of the SH⁻ fragment would also be spread from zero energy. Also, it may not show the anisotropy obtained in the momentum image. Hence, we conclude that this concerted break-up mechanism is contributing very little to the signal. At 8 eV electron energy, in path (5), the excess energy available to the system is 7.15 eV. However, the $C_2H_5$ radical can also be electronically excited. Two electronic excited states of this radical are identified in the UV absorption spectra at 5.03 and 6.05 eV [35,36]. This would provide the following additional dissociation paths for the resonance state:

$$e + C_2H_5SH \rightarrow (C_2H_5SH^-)^* \rightarrow (C_2H_5)\,(3s) + SH^- \quad (7)$$

$$e + C_2H_5SH \rightarrow (C_2H_5SH^-)^* \rightarrow (C_2H_5)\,(3p) + SH^- \quad (8)$$

The corresponding excess energy available in the system would be 2.12 and 1.1 eV, respectively. This implies that the maximum kinetic energy observable in the SH⁻ channel would be 0.99 and 0.52 eV, respectively. As the observed kinetic energy distribution at 8 eV extends up to 1.3 eV, considering the electron energy resolution, we attribute the observed SH⁻ channel to path (7).

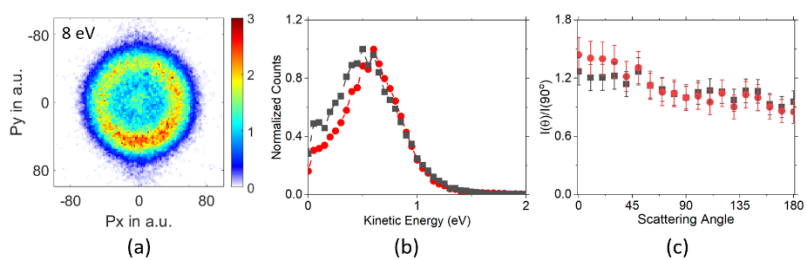

Figure 7: (a) Momentum image obtained for the S⁻/SH⁻ channel from and 1-butanethiol at 8 eV. Corresponding (b) kinetic energy and (c) angular distribution of S⁻/SH⁻ channel from 1-butanethiol (-■-) along with those for ethanethiol (-●-).

The momentum image obtained at 8 eV for the 1-butanethiol also shows a ring similar to that in ethanethiol (Figure 7). Similar excited states are also observed in secondary butyl radical[35]. We assume that the primary butyl radical would also have similar states and attribute this channel in 1-butanethiol to such electronically excited butyl radical similar to ethanethiol. 1-butanethiol also shows comparatively lower anisotropy, indicating the role of a higher number of degrees of freedom in delaying the initial dissociation process.

At 9 eV, the ion signal is dominated by the $SH^-$ ions (Figure 5) and the corresponding momentum image shows a substantial increase in the low kinetic energy ions (Figure 5(c) and 6(a)). This indicates that the contribution from path (8) starts showing up with significant intensity. It also shows that this particular path also results in internal excitation of the $C_2H_5$ radical. Moreover, path (6) may also have a considerable contribution to the low energy ion signal at this energy.

### 3.3 $(M-1)^-$ ions:

This channel shows a peak at 2.2 eV in ethanethiol and 1.8 eV in 1-butanethiol, and the peak cross section of this channel is $7.70 \times 10^{-20}$ and $7.56 \times 10^{-20}$ cm², respectively. In ethanethiol, the $(M-2)^-$ ions are also observed at lower energies[12]. However, their signal is found to be extremely low at these energies. Hence, although we cannot distinguish between the two ions due to the limited mass resolution of our mass spectrometer, we attribute the observed signal at the peak to the $(M-1)^-$ ions. This channel is attributed to the single particle shape resonance based on its appearance energy. We have obtained the momentum image of this channel. The image shows a slightly elongated blob along the electron beam with very little momentum. This corresponds to the kinetic energy ranging up to 50 meV. Considering the threshold of formation of this channel as 1.74 eV[12], the maximum kinetic energy expected in this channel would be 10 meV. Thus, the observed spread in the image may be due to the combination of the poorer electron energy resolution, thermal spread of the target molecules and the imperfect imaging condition at such a low kinetic energy.

**Conclusion:**

We have observed the thiol functional group-dependent site selectivity in DEA. The dynamics of the $H^-$ channel resemble that for hydrogen sulphide, the precursor molecule for this functional group. Their resonance positions match fairly well. At 5 eV, the angular distribution also partially resembles that for the hydrogen sulphide. However, like ethanol, this channel shows the effect of torsional motion. This effect does not change with the number of degrees of freedom, as seen in the comparison between ethanethiol and 1-butanethiol. The 7 eV resonance shows an angular distribution similar to hydrogen sulphide. The kinetic energy distribution shows that very little energy is gone to the molecular fragment's internal degrees of freedom, showing dissociation dynamics following the axial recoil motion. The third resonance at 8 eV clearly shows two contributions: the low energy signal that corresponds to the C-H site break, which is known to be a many-body breakage mechanism related to the alkyl site, and the higher energy channel with anisotropic angular distribution with the alkanthio radical in its electronic excited state. The $S^-/SH^-$ channel shows a single peak at around 8 eV. The momentum images across this peak show two distinct contributions. The $S^-$ channel contributes substantially in the rising edge of the peak and shows the momentum image as a ring. We attribute this channel to the sequential dissociation of electronically excited $SH^-$ ion. At 8 eV both $S^-$ and $SH^-$ channels contribute. The corresponding $SH^-$ signal in the ring is understood to be from the formation of electronically excited $C_2H_5$ ($3s$) radical and stable $SH^-$ ion. On the trailing edge of the peak, the $SH^-$ ion signal dominates and is attributed to the formation of electronically excited $C_2H_5$ ($3p$) radical. The $(M-1)^-$ channel shows very little kinetic energy, as expected due to the proximity of the resonance to the threshold of this channel. The overall molecular dynamics observed in ethanethiol and 1-butanethiol for various DEA channels show marked similarity and these features can be generalised for the alhylthiols.

## Acknowledgement

The authors acknowledge the financial support from the Department of Atomic Energy, India, under Project Identification No. RTI4002.